\documentclass[a4paper,11pt]{article}
\pdfoutput=1
\usepackage{amsmath,amsthm,latexsym,amssymb,amsfonts,epsfig}
\usepackage{fontenc,dsfont,layout}
\usepackage{hyperref} 
\usepackage{psfrag}
\usepackage[cp1250]{inputenc}

\usepackage{graphicx}

\usepackage{epstopdf} 
\usepackage{caption}
\usepackage{float}
\usepackage{multirow} 
\usepackage{rotating}
\usepackage{pbox}
 
\numberwithin{equation}{section}

\DeclareMathAlphabet{\mathpzc}{OT1}{pzc}{m}{it}

\begin{document}

\title{Higher-order  scalar interactions and SM vacuum stability}
\author{Zygmunt Lalak\footnote{Zygmunt.Lalak@fuw.edu.pl}{}\;  Marek Lewicki\footnote{Marek.Lewicki@fuw.edu.pl}{}\;  and Pawe\l{}  Olszewski\footnote{Pawel.Olszewski@fuw.edu.pl} \\ 
Institute of
Theoretical Physics, Faculty of Physics, University of Warsaw\\ ul. Ho\.za 69,
Warsaw, Poland} 
\date{}
\maketitle

\begin{abstract}
Investigation of the structure of the Standard Model effective potential at very large field strengths opens a window towards new phenomena and can reveal properties of the UV completion of the SM. The map of the lifetimes of the vacua of the SM enhanced by nonrenormalizable scalar couplings has been compiled to show how new interactions modify stability of the electroweak vacuum. Whereas it is possible to stabilize the SM by adding Planck scale suppressed interactions and taking into account running of the new couplings, the generic effect is shortening the lifetime and hence further destabilisation of the SM electroweak vacuum. These findings have been illustrated with  phase diagrams of modified SM-like models. It has been demonstrated that stabilisation can be achieved by lowering the suppression scale of higher order operators while picking up such combinations of new couplings, which do not deepen the new minima of the potential. Our results show the dependence of the lifetime of the electroweak minimum on the magnitude of the new couplings, including cases with very small couplings (which means very large effective suppression scale) and couplings 
vastly different in magnitude (which corresponds to two different suppression scales). 
\end{abstract}

\section{Introduction}
The discovery of the 126 GeV scalar particle, which in the light of available data can be identified with the Standard Model Higgs boson,  and absence of   experimental signature of any new physical state in the LHC experiments makes it important to search for possible windows towards new phenomena within the Stadard Model itself. One of the possible windows is the investigation of the structure of the effective potential in the Standard Model which has been the subject of considerable activity \cite{Branchina:2013jra} \cite{Buttazzo:2013uya}  \cite{Degrassi} \cite{Ellis:2009tp} \cite{Espinosa} \cite{Casas:2000mn} \cite{Casas:1996aq} \cite{Casas:1994qy} \cite{Sher}. 

The study of the renormalisation group improved effective SM  potential has revealed an interesting structure at field strenghts  higher than approximately $10^{11}$ GeV and new minima at superplanckian field strenghts. The upshot depends critically on the precise value of the measured Higgs mass and on the measured value of the top quark Yukawa coupling. 
In particular, one finds that for the central value of the top mass and for the central value of the measured Higgs mass the physical electroweak symmetry breaking minimum becomes metastable with respect to the tunneling from the physical EWSB minimum to a deeper minimum located at superplanckian values of the Higgs field strength. The computed lifetime of the metastable SM Universe turns out larger than the presently estimated age of the Universe, however  the instability border in the space of parameters $M_{top} - M_{higgs}$ looks uncomfortably close and this suggests that the result is rather sensitive to various types of modifications that can be brought in by the BSM extensions. 

The question about stability of the SM vacuum in the presence of ultraviolet completions at or below the planck scale is the central point of this note.  In the paper \cite{Branchina:2013jra} higher order operators have been added to the scalar potential of the neutral higgs field. The operators are suppressed by suitable powers of the Planck scale and for sensible values of the new couplings they were found to modify significantly the behaviour of the potential near the Planck scale. It has been shown and illustrated by  examples in \cite{Branchina:2013jra}, that Planck scale operators can indeed produce a minimum near the Planck scale, however they can also destabilize the SM metastable minimum with respect to the tunneling to a deeper minimum  at high field strenghths. 

Here we study the question further, making a more complete map of the vacua in the SM extended by nonrenormalisable scalar couplings. taking into account the running of the new couplings and going beyond the standard assumptions taken when calculating the lifetime of the metastable vacuum. Usually, one uses certain quasi-analytic approximations of the effective potential, \cite{WeinbergLee}, for the purpose of calculating the tunneling rate. In particular, one uses as the departure point the calculation for the quartic-like form of the effective potential at large field strenghts, while in the modified scalar potential it is the order 6 or order 8 coupling which naively dominates the potential in the large field domain. It is important to check the validity of such approximation and to search through a relatively wide scope of new couplings to find the actual behaviour of the scalar potential. The ultimate tool in this case is the direct numerical analysis, which however is not so straightforward because of the flatness of the effective potential. In this paper we present currently available resulats of such extended analysis of the modified SM scalar potential. For the purpose of the present paper we have suppressed nonrenormalisable operators with derivatives. In general, we confirm that it is relatively easy to destabilize the  SM with the help of the Planck scale suppressed scalar operators. However, there exists  choices of higher-dimensional couplings which meta-stabilize the SM vacuum.  

\section{New interactions}
\begin{figure}[h!]
\begin{minipage}[t]{0.5\linewidth} 
\centering
\includegraphics[scale=0.87]{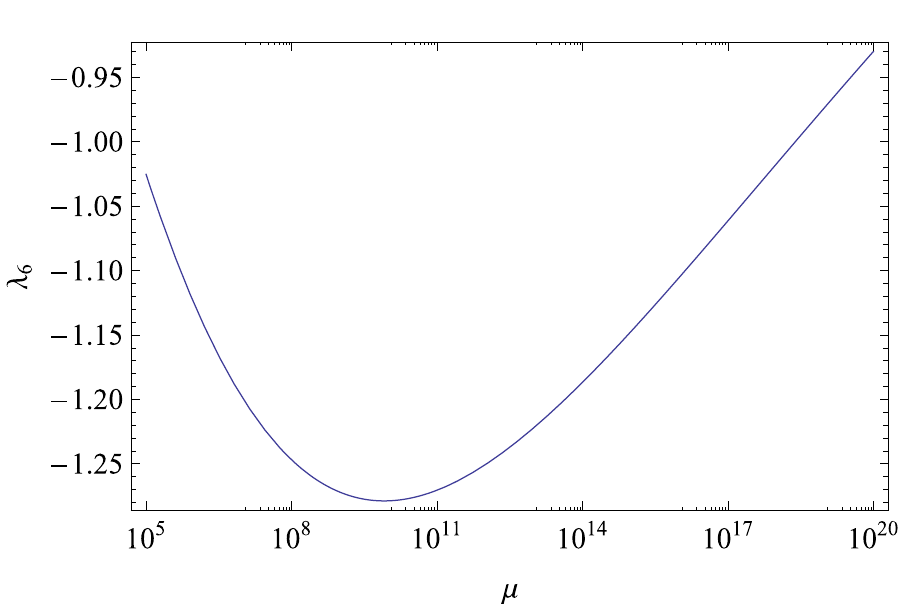} 
\end{minipage}
\hspace{0.5cm}
\begin{minipage}[t]{0.5\linewidth}
\centering 
\includegraphics[scale=0.87]{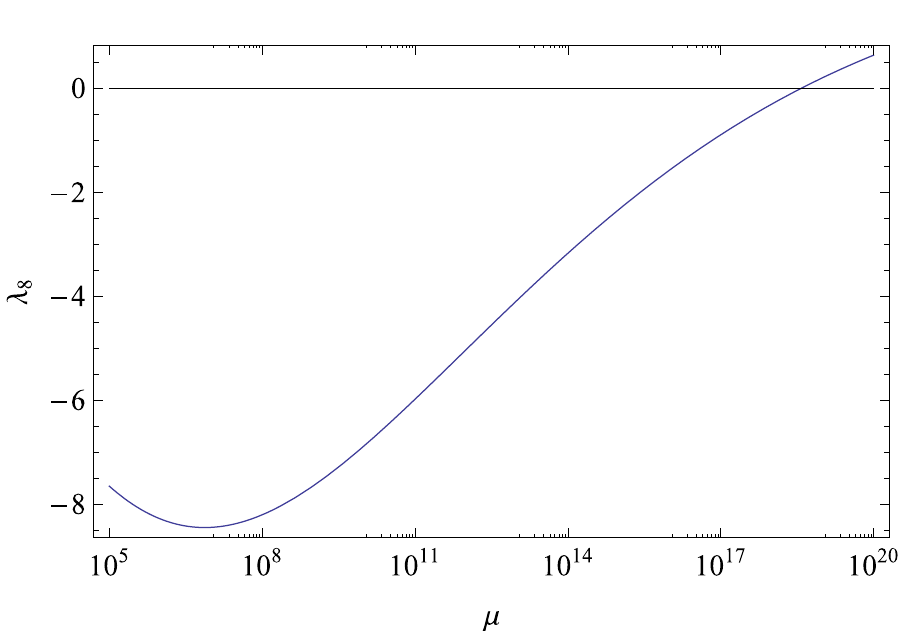}  
\end{minipage}
\captionof{figure}{
Example solution of renormalisation group equations for couplings $\lambda_6(M_p)=-1$ and $\lambda_8(M_p)=-0.1$. 
\label{running}
}
\end{figure}

\begin{figure}[h!]
\begin{minipage}[t]{0.5\linewidth} 
\centering
\includegraphics[scale=0.87]{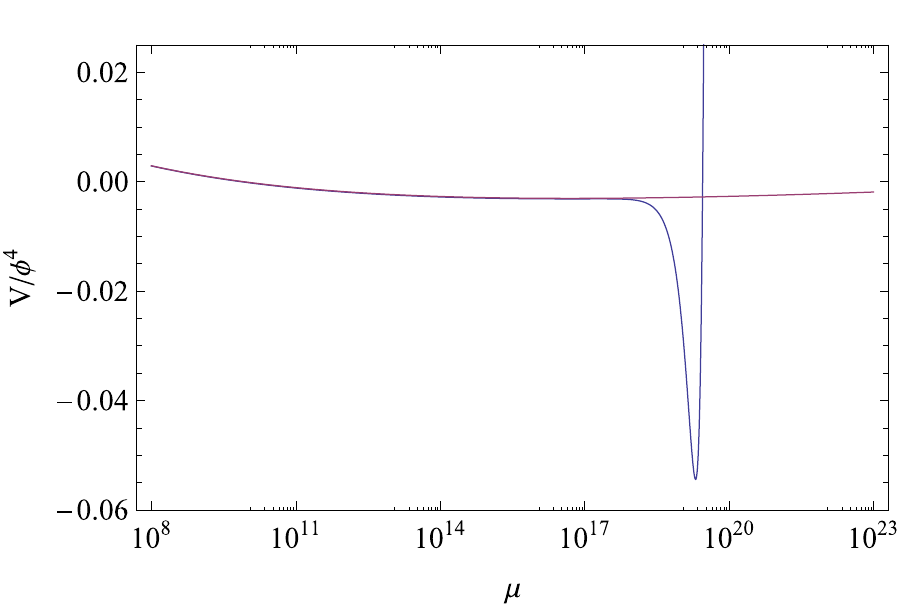} 
\end{minipage}
\hspace{0.5cm}
\begin{minipage}[t]{0.5\linewidth}
\centering 
\includegraphics[scale=0.87]{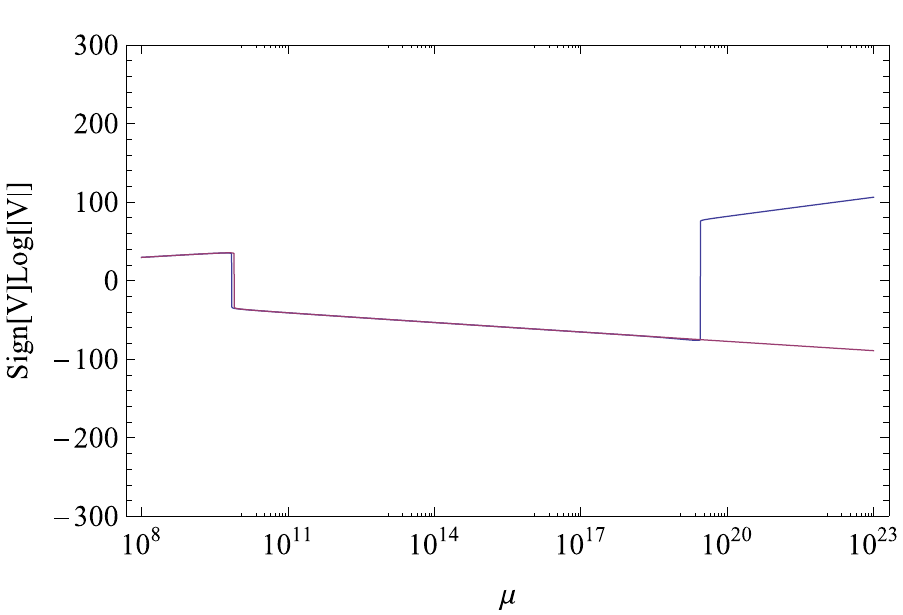}  
\end{minipage}
\captionof{figure}{
Potential corresponding to couplings from Figure~\ref{running}~(blue line) together with the Standard Model potential (purple line).
\label{potentialplot}
}
\end{figure}

In what follows we shall assume the Lagrangian of the Standard Mode augmented by two higher dimensional operators proportional to $|H|^6$ and $|H|^8$, where $H$ is the Higgs doublet. They are suppressed by a large mass scale $M$ to an appropriate power. Being interested only in the direction $H=(\phi / \sqrt{2},0)$, we obtain a potential of the form (similar to \cite{Branchina:2013jra}):
\begin{equation}\label{potential}
V=-\frac{m^2}{2} \phi^2 +\frac{\lambda}{4} \phi^4 + \frac{\lambda_6}{6!}\frac{\phi^6}{ M^2} + \frac{\lambda_8}{8!}  \frac{\phi^8}{M^4}.
\end{equation}
It is well known that effects of radiative corrections to SM couplings on the vacuum are large and so these couplings require precise determination \cite{Buttazzo:2013uya}. We have used two-loop running of the SM parameters \cite{Espinosa} and computed one-loop corrections to the new interactions. The correction to the running of the quatric Higgs coupling is of the form 
\begin{equation}\label{dlambda}
\Delta \beta_{\lambda}=\frac{\lambda_6}{16\pi^2}\frac{m^2}{M^2},
\end{equation}
and its contribution is negligible for $m<<M$. One-loop beta functions of new couplings take the form
\begin{eqnarray} \label{betafunctions}
16\pi^2\beta_{\lambda_6} &=& \frac{10}{7}\lambda_8\frac{m^2}{M^2}+18\lambda_6 6\lambda-6\lambda_6\left(\frac{9}{4}g_2^2+\frac{9}{20}g_1^2-3y_t^2\right), \\ 
16\pi^2\beta_{\lambda_8} &=& \frac{7}{5}28\lambda_6^2+30\lambda_8 6\lambda-8\lambda_8\left(\frac{9}{4}g_2^2+\frac{9}{20}g_1^2-3y_t^2\right), \nonumber
\end{eqnarray}
which agrees with \cite{Jenkins:2013zja}.
Figure~\ref{running} shows an example of running of the new couplings and Figure~\ref{potentialplot}  shows the resulting potential with 
$\lambda_6(M_p)=-1$, $\lambda_8(M_p)=-0.1$ and suppression scale $M=M_p$.

\section{Tunneling rate}
To calculate the expected lifetime of a metastable vacuum present in potential $V(\phi)$ we use the standard formalism of finding a bounce solution \cite{Coleman} \cite{Callan} which in the $O(4)$ symmetric case depends only on $s=\sqrt{\vec{x}^2+x_4^2}$. This means solving an equation of motion of the form 
\begin{equation}
\ddot\phi + \frac{3}{s}\dot\phi=\frac{\partial V(\phi)}{\partial \phi},\label{phi_eqn}
\end{equation}
with a dot denoting a derivative with respect to $s$.  The boundary conditions are $\dot\phi(0)=0$, so that the solution is non-singular at $s=0$, and $\phi(\infty)=\phi_{\textrm{min}}$ so that it corresponds to the decay of the metastable vacuum positioned at $\phi_\textrm{min}$.

The above assumes canonical form of the kinetic term $(\partial \phi)^2$ in the Lagrangian. We have suppressed nonrenormalisable operators with derivatives. For operators with two derivatives one can use field redefinition suggested in \cite{Isidori:2007vm}:  $\phi\longrightarrow  \tilde{\phi} \left(1+a \frac{\tilde{\phi}^2}{M^2}+b \frac{\tilde{\phi}^4}{M^4} \right)$, to bring  troublesome operators   
$\partial_\mu \phi \partial^\mu \phi \left(c_1 \frac{\phi^2}{M^2}+c_2 \frac{\phi^4}{M^4}\right)$ to the form of the operators already included in \eqref{potential} which allows us to use \eqref{phi_eqn}  with shifted couplings $\lambda_6 \rightarrow \tilde{\lambda}_6$ and $\lambda_8 \rightarrow \tilde{\lambda}_8$. However, derivative operators which are not of this form, like operators with 4 derivatives, induce a modification of  \eqref{phi_eqn} that can further destabilize the already unstable vacua. Analysis of the complete set of operators of order eight  lies beyond the scope of this paper.


Having found the bounce, we calculate its euclidean action given by  
\begin{equation}
S_E=\int d^4 x\left\{ \frac{1}{2}\sum_{\alpha=1}^4\left(\frac{\partial \phi(\bf{x})}{\partial x^\alpha}\right)^2 +V(\phi(\bf{x})) \right\}
=2\pi^2\int ds s^3\left( \frac{1}{2} \dot\phi^2(s) +V(\phi(s)) \right),
\end{equation}
which allows us to calculate decay probability of a volume $d^3x$ 
\begin{equation}
dp=dtd^3x\frac{S_E^2}{4 \pi^2}\left| \frac{det'[-\partial^2+V''(\phi)]}{det[-\partial^2+V''(\phi_{\textrm{min}})]} \right|^{-1/2}e^{-S_E}.
\end{equation}
To calculate the expected lifetime we simply integrate that probability assuming size of the universe $T_U=10^{10}\textrm{yr}$ in the spatial directions and define the expected lifetime $\tau$ as time at which decay probability is equal to 1. We also approximate the determinant and normalization prefactor  by another dimensionfull quantity encountered in our problem, namely $\phi_0=\phi(0)$. The error introduced that way is small compared to uncertainty in determination of action, because lifetime depends only on fourth power of $\phi_0$ while its dependence on action is exponential,
\begin{equation}\label{lifetime}
\frac{\tau}{T_U}=\frac{1}{\phi_0^4 T_U^4}e^{S_E}.
\end{equation}
In the following sections we will present known analytical approximations and compare their results with our numerical solution. As the suppression scale we use the Planck mass, $M=M_P$, unless stated otherwise (see: the \textit{Lowering the magnitude of the suppression scale} section below). 

\subsection{Analytical solution}\label{anatyticalresults}
Possibly the simplest scheme for estimating the vacuums lifetime, often conjectured for the SM (eg. \cite{Buttazzo:2013uya}), amounts to calculating the quantity in \eqref{lifetime} as
\begin{equation}
\frac{\tau}{T_U}=\frac{1}{\Lambda_B^4 T_U^4}e^{\frac{8\,\pi^2}{3 \,}\frac{1}{|\lambda_{eff}(\Lambda_B)|}}\,,
\label{simpletime}
\end{equation}
where $\frac{\lambda_{eff}(\phi)}{4} = \frac{V_{eff}(\phi)}{\phi^4}$, $V_{eff}$ being the effective potential, and $\Lambda_B$ denoting a renormalisation scale that minimises $\lambda_{eff}$.
This approach utilises the fact that for a wide range of energy scales, $\lambda_{eff}$ is close to a constant negative value $-|b|\approx -0.014$. Bounce solution for the simple quartic potential of the form $-\frac{|b|}{4}\phi^4$ is known \cite{WeinbergLee} and its action is exactly $\frac{8\, \pi^2}{3\,}\frac{1}{|b|}$. Thus, taking minimum of $\lambda_{eff}$, one estimates the action of a true bounce from below. Simultaneously, the picked $\Lambda_B$ value serves as the only characteristic scale for the bounce since the quartic potential is classically scale invariant. 

Simplifying further, we approximate $\lambda_{eff}$ only by the RGE-improved quasiquartic coupling
$\lambda_4(\phi)+\frac{4}{6!}\frac{\lambda_6}{M_P^2}\phi^2 + \frac{4}{8!}\frac{\lambda_8}{M_P^4}\phi^4$, neglecting both the mass term and running of the field itself in the tree-level $V_{eff}$. We use two approaches. Firstly we also completely ignore the RGE-running of $\lambda_6$ and $\lambda_8$. Then we include them in the set of RGE equations and make them scale dependent according to \eqref{dlambda} and \eqref{betafunctions}.

\begin{figure}[H]
\begin{minipage}[t]{0.5\linewidth} 
\centering
\includegraphics[scale=0.87]{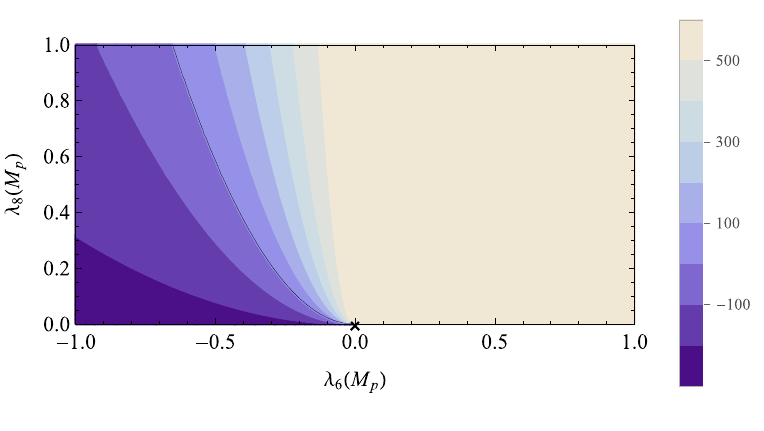} 
\end{minipage}
\hspace{0.5cm}
\begin{minipage}[t]{0.5\linewidth}
\centering 
\includegraphics[scale=0.87]{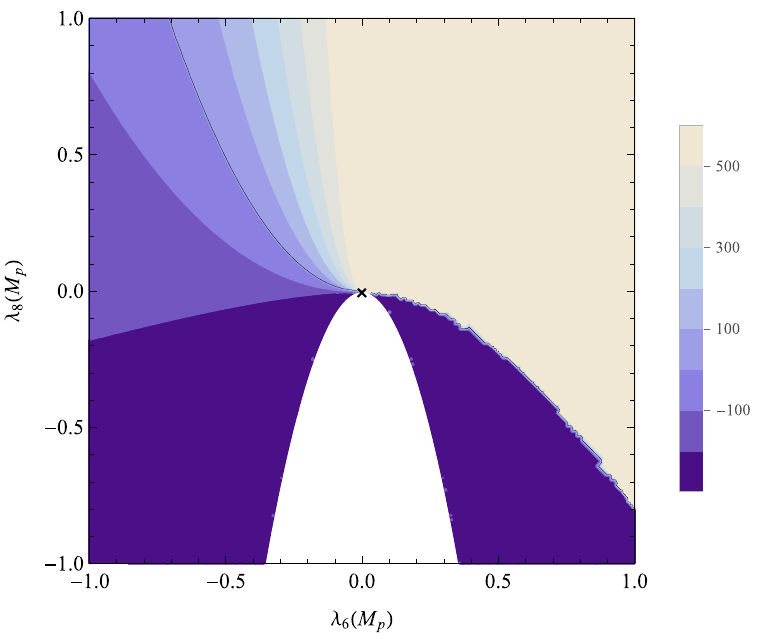}  
\end{minipage}
\captionof{figure}{Decimal logatihm of lifetime of the universe in units of $T_U$ as a function of the nonrenormalisable $\lambda_6$ and $\lambda_8$ couplings, calculated with formula (\ref{simpletime}). For $\lambda_6$ and $\lambda_8$ kept constant (left panel) and $\lambda_6$ and $\lambda_8$ scale dependent and satisfying their one-loop RGE's (right panel).} 
\label{pseudoanalitic2}
\end{figure}

When $\lambda_6$ and $\lambda_8$ do not run, $\lambda_{eff}$ does not have a global minimum for $\lambda_8 < 0$. Thus we can calculate the value \eqref{simpletime} only for the range of positive $\lambda_8$'s. The lhs of Figure \ref{pseudoanalitic2} shows a contour plot of $\log_{10} \frac{\tau}{T_U}$ for $-1<\lambda_6<1$ and $0<\lambda_8<1$. In the region where $\lambda_6$ is negative enough, $\lambda_{eff}$ develops new minimum (as compared to SM) at scales close to $M_P$ and the exponent in (\ref{simpletime})  becomes small, rendering the vacuum short-lived.

Next we include the running of $\lambda_6$ and $\lambda_8$. It has small influence on the position of the $\log_{10} \frac{\tau}{T_U}=0$ contour. The novelty is that now, even when put negative at the Planck scale, $\lambda_8$ eventually becomes positive and so $\lambda_{eff}$ posseses global minimum, thus enabling us to use the formula (\ref{simpletime}) in wider range of $\lambda_8$'s. The right panel of Figure \ref{pseudoanalitic2}. shows the plot of $\log_{10} \frac{\tau}{T_U}$ for the values of $\lambda_6(M_P)$ and $\lambda_8(M_P)$ between $-1$ and $1$ as put at the Planck scale. The region where $\lambda_{eff}$ does not develop a global minimum at the renormalisation scale lower than $M_P^2$ was excluded (white color).

Another way of breaking the scale invariance of quartic potential (originally presented in \cite{WeinbergLee} and recently used in \cite{Branchina:2013jra}) would be to sew it with a linear function 
\begin{equation}\label{approxV}
V_{\eta}(\phi) =
  \begin{cases}
   -\frac{b_\eta}{4} \phi^4 \;,& \phi \leqslant \eta \\
   -\frac{b_\eta}{4} \eta^4 - K\,(\phi-\eta)  \;,&   \phi\ > \eta
  \end{cases}\;,
\end{equation}
where $\frac{b_\eta}{4}=\frac{-V_{eff}(\eta)}{\eta^4}=\frac{-\lambda_{eff}(\eta)}{4}$.
 One then still needs to choose the sewing point $\eta$ and the slope parameter $K$, to approximate the effective potential.
It is conceivable to sew the two functions at any point, on the plot of $V_{eff}$ on left hand side of Figure~\ref{potentialplot}. 
Evidently certain level of arbitrariness is present in choosing the bounce this way.

Moreover, no special bounce solution is singled out unless the ratio of derivatives at $\eta$, $-\gamma=\frac{b_\eta\,\eta^3}{K}$, falls into the region $0<-\gamma<1$, \cite{WeinbergLee}. The action of a bounce is then given by $\displaystyle S_{\eta}= \frac{8\pi^2}{3 \,} \frac{1}{b_\eta} (1-(\gamma+1)^4)$, and its starting point, $\phi_0=\eta(2+\gamma)$, lies in the linear part of $V_\eta$. In the case when $-\gamma \geqslant 1$, all the bounce solutions lie in the quartic part of $V_\eta$ only and the effect of sewing it with the linear function amounts again to arbitrarily choosing the value of quartic coupling $\lambda_{eff}(\eta)$ (and the scale of the bounce $\Lambda_B=\eta$) in \eqref{simpletime}.

One could be tempted to destabilise the vacuum by a steep linear function (large $K$) but generally, in order for $V_\eta$ to seemingly reproduce the shape of $V_{eff}$ near the global minimum, one has to put $-\gamma$ close to $1$ and $\eta$ of the order of $M_P$. It follows that the main effect of lowering the action in \eqref{lifetime} by the unrenormalisable operators comes from the increase of $|\lambda_{eff}|$, just like in the previously described scheme.

As a check, we have calculated the $\log_{10} \frac{\tau}{T_U}$ (for nonrunning $-1<\lambda_6<1$, $0<\lambda_8<1$), putting $-K=V_{eff}'(\eta)$ and $\eta$ equal to the potentials inflection point ($V_{eff}''(\eta)=0$). The results were qualitatively similar to the ones presented above: in particular the $\log_{10} \frac{\tau}{T_U}=0$ contour remained practically unmoved.
\subsection{Numerical result}
The numerical procedure we used is based on finding solutions to the equation of motion \eqref{phi_eqn}, by an overshot/undershot method. First we solve starting from the true vaccuum at very small $s=\epsilon$ and expanding the solution into a series to get
\begin{eqnarray}
\phi &\approx & \phi_0 + \frac{\epsilon^2}{8}\left. \frac{\partial V(\phi)}{\partial \phi}\right|_{\phi=\phi_0}, \\ 
\dot{\phi} &\approx &\frac{\epsilon}{4} \left. \frac{\partial V(\phi)}{\partial \phi}\right|_{\phi=\phi_0}. \nonumber
\end{eqnarray}
Than we use simple bisection to find $\phi_0$ for which 
$\phi(\infty)$ is the electroweak minimum.

 Next we solve the equation of motion again, this time starting from the electroweak minimum. We first expand the field and potential around the minimum
\begin{eqnarray}
\phi &\approx & \phi_\textrm{min}+\phi_\infty , \\ 
\frac{\partial V(\phi)}{\partial \phi} &\approx & m^2\phi_\infty . \nonumber
\end{eqnarray}
Thus we get a simplified equation of motion which is solved by modified Bessel functions, so we can express the initial conditions as
\begin{eqnarray}
\phi_\infty &= & A \ \frac{K_1(s)}{s}, \\ 
\dot{\phi}_\infty &= & -A \ \frac{K_2(s)}{s}.
\end{eqnarray}
These conditions are  solved to obtain $\dot{\phi}_\infty$ as a function of $\phi_\infty$. We then again use simple bisection to find $\phi_\infty$ which minimizes the field derivative at a very small $s=\epsilon$ near the true vacuum. 

Numerical problems arise in this scenario because standard model potential is very flat and the change induced by the new couplings appears only around  the Planck scale, so we have to solve the equation of motion through sixteen orders of magnitude in the field 
$\phi$. For this reason it is very hard to choose numerical values of $\epsilon$ and $\infty$ for parameter $s$ such that the bisection converges to the desired solution. Hence we only required that one of the above methods converged at any given point.  
 
\begin{figure}[H]
\centering
\includegraphics[scale=1.2]{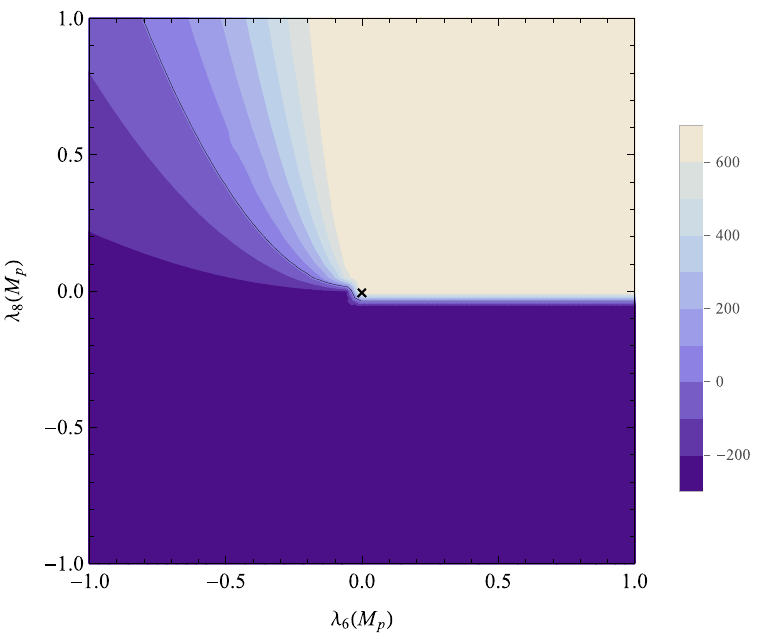} 
\captionof{figure}{
Decimal logarithm of lifetime of the universe in units of $T_U$ as a function of constant couplings $\lambda_6$ and $\lambda_8$. 
\label{constl}
}
\end{figure}

The resulting lifetimes are shown in Figure~\ref{constl} for constant couplings $\lambda_6$ and $\lambda_8$. 
We can distinguish 3 different areas in Figure~\ref{constl}. The first one with both new couplings positive corresponds to the SM potential stabilised by new interactions at the Planck scale. The resulting lifetimes are very close to SM one, because the bounce solution with SM potential starts at field values smaller than Planck mass ($\phi_0<M_p$). The second region with $\lambda_8<0$ corresponds to a potential unbounded from below, and as we can see a quickly decaying bounce solution appears when $\lambda_8$ becomes negative. The last region with positive $\lambda_8$ but negative $\lambda_6$ corresponds to a stabilized potential with a new minimum around the Planck scale which can be approximated with an analytical solution described in the previous section. 

\begin{figure}[H]
\centering
\includegraphics[scale=1.2]{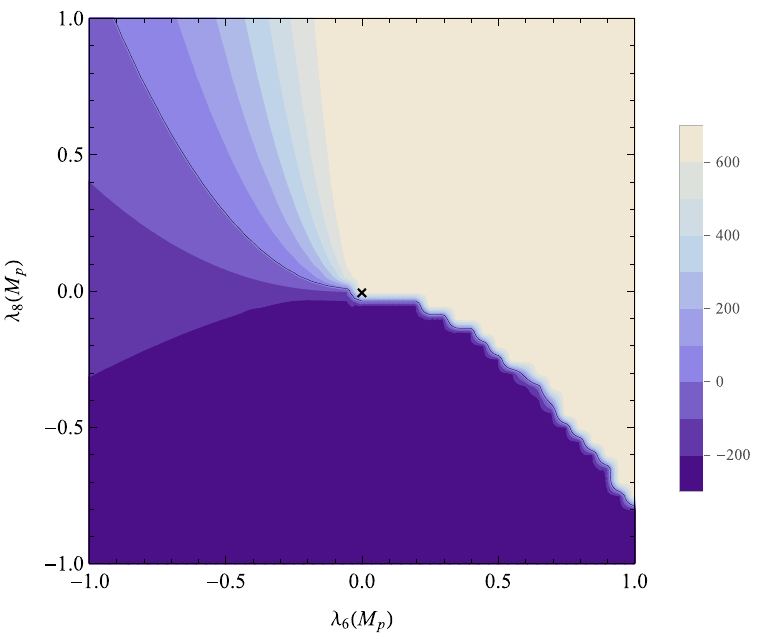} 
\captionof{figure}{
Decimal logarithm of lifetime of the universe in units of $T_U$ as a function of running couplings $\lambda_6$ and $\lambda_8$ calculated at the scale $M$. 
\label{runningl}
}
\end{figure}

To further increase the accuracy of above prediction we solved the equation of motion \eqref{phi_eqn} numerically taking into account the 1-loop running of  $\lambda_6$ and $\lambda_8$ from equation (2.3) together with 2-loop Standard Model RGEs. The resulting lifetimes are shown in Figure~\ref{runningl}.

Examples of running of new couplings in Figure~\ref{running} show that their values can change significantly, but the most important qualitative difference comes from the $\lambda^2_6$ contribution to the running of $\lambda_8$ (see equation \eqref{betafunctions}). In Figure~\ref{running} we see that for large enough $\lambda_6$ it can stabilize the potential by pushing $\lambda_8$ to positive values not far above the Planck scale, when $\lambda_8(M_p)$ is negative but has small enough modulus. This effect bends the metastability curve in Figure~\ref{runningl} towards more negative $\lambda_8$ near edges of the plot where $|\lambda_6|$ is large.
\subsection{Comparison of different methods}
\begin{figure}[H]
\centering
\includegraphics[scale=1.3]{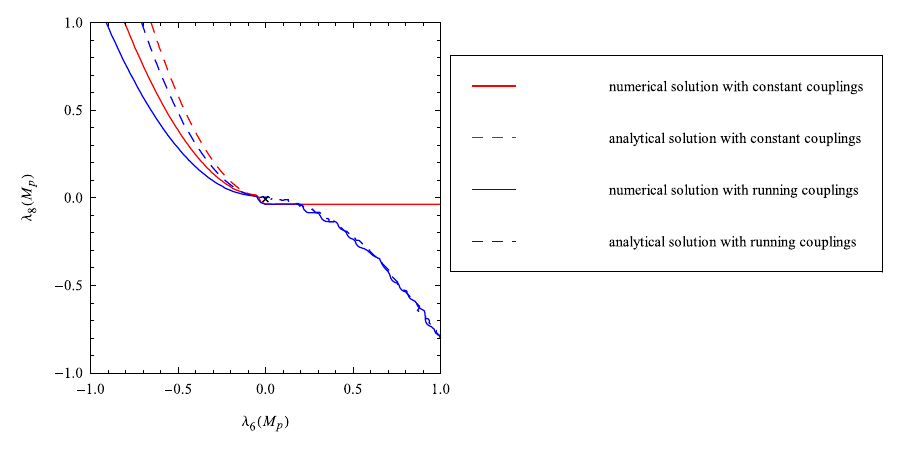} 
\captionof{figure}{
Contours corresponding to metastability boundary ($\tau=T_u$) obtained using four different methods. 
\label{comparel}
}
\end{figure}

Figure~\ref{comparel} shows comparison of results obtained using the methods described above. The analytical approximation is accurate enough for qualitative analysis, however more careful numerical analysis results in a larger stability region. The same can be said about the effect of taking into account running of nonrenormalisable couplings, where sufficiently large contribution from $\lambda_6$ to running of 
$\lambda_8$ can save otherwise unstable vaccua.  
\section{Standard Model  phase diagram}
\begin{figure}[H]
\begin{minipage}[t]{0.31\linewidth} 
\centering
\includegraphics[scale=0.65]{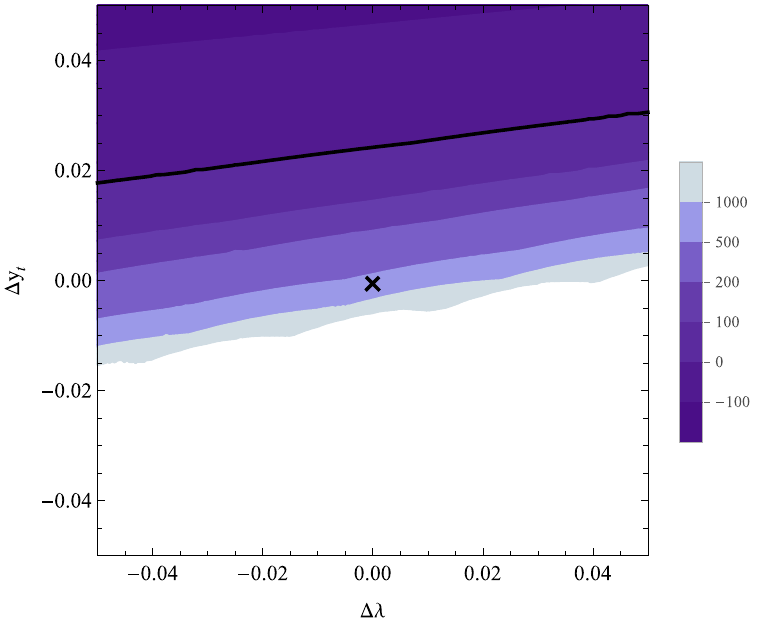} 
\end{minipage}
\hspace{0.5cm}
\begin{minipage}[t]{0.32\linewidth}
\centering 
\includegraphics[scale=0.65]{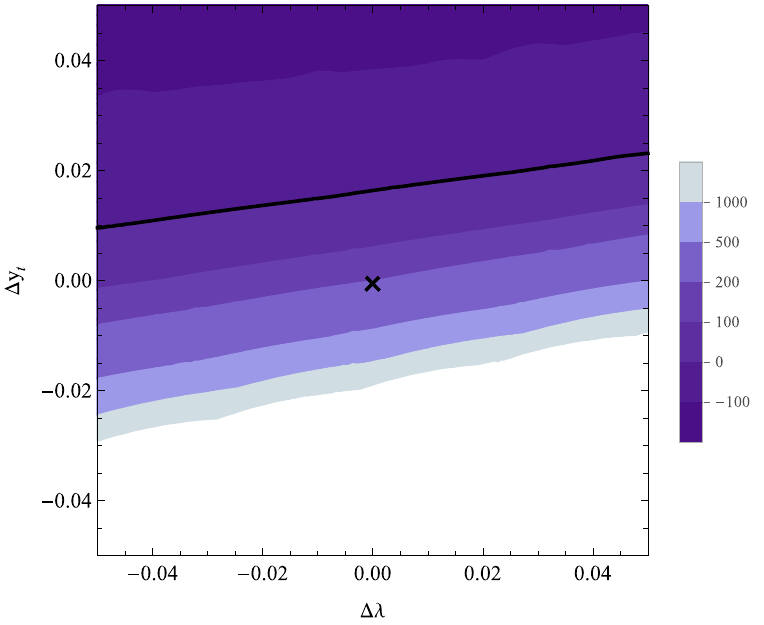}  
\end{minipage}
\begin{minipage}[t]{0.31\linewidth}
\centering 
\includegraphics[scale=0.65]{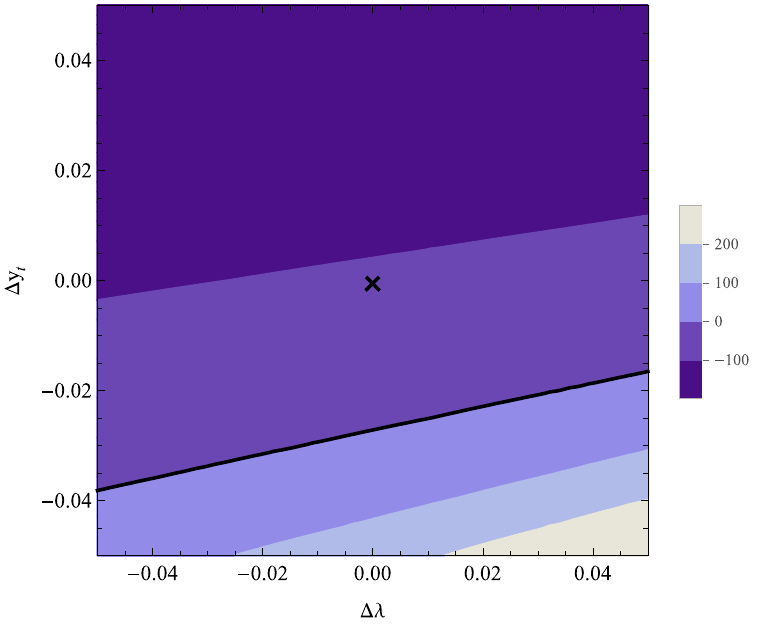}  
\end{minipage}
\captionof{figure}{
Standard Model phase diagram (left panel), the same diagram after inluding new operators $\lambda_6(M_p)=-1/2$ and  $\lambda_8(M_p)=1$ (middle panel) and $\lambda_6(M_p)=-1$ and  $\lambda_8(M_p)=1/2$ (right panel). The white region corresponds to absolute stability, and the black line marks the instability border.
\label{phasediagram}
}
\end{figure}

To illustrate effects of new nonrenormalisable operators on Standard model vacuum stability in
Figure~\ref{phasediagram} we show the well known standard model phase diagram (see for example \cite{Buttazzo:2013uya}) and the same diagram after inluding new operators, respectively  $\lambda_6(M_p)=-1/2, -1$ and  $\lambda_8(M_p)=1, 1/2$.
Above plots clearly show that nonrenormalisable interactions supressed  by the Planck mass can drastically change the SM phase diagram, by pushing electroweak vacuum towards the instability region. 

\section{Lowering the magnitude of the suppression scale}
\begin{figure}[H]
\centering
\includegraphics[scale=1.4]{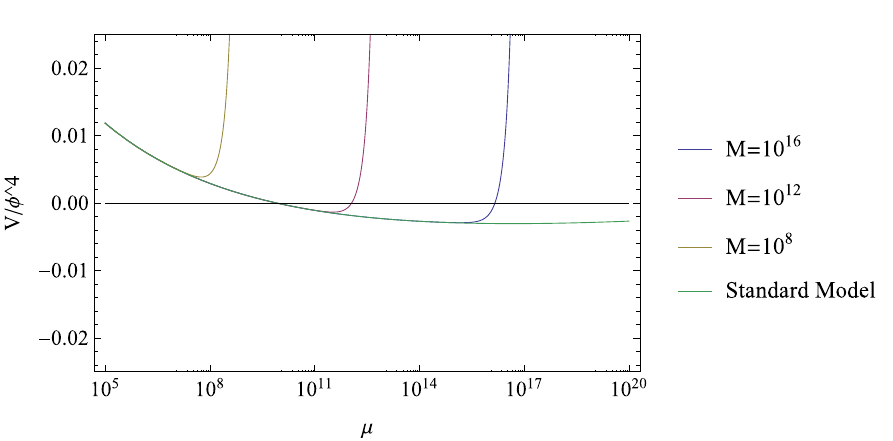} 
\captionof{figure}{
Scale dependence of $\frac{\lambda_{eff}}{4}=\frac{V}{\phi^4}$ with $\lambda_6=\lambda_8=1$ for different values of suppression scale $M$. The lifetimes corresponding to suppression scales $M=10^{8},10^{12},10^{16}$ are, respectively, $\log_{10}(\frac{\tau}{T_U})=\infty,1302,581$ while for the Standard Model $\log_{10}(\frac{\tau}{T_U})=540$.
\label{scaleplots1}
}
\end{figure}

\begin{figure}[H]
\centering
\includegraphics[scale=1.4]{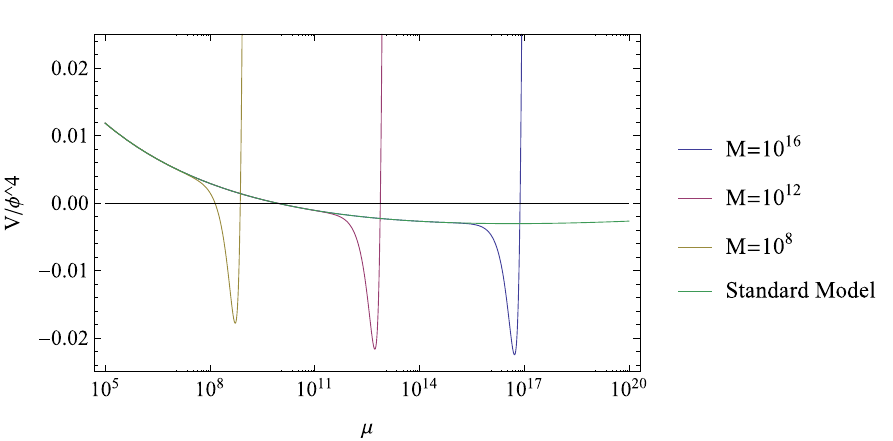} 
\captionof{figure}{
Scale dependence of $\frac{\lambda_{eff}}{4}=\frac{V}{\phi^4}$ with $\lambda_6=-1$ and $\lambda_8=1$ for different values of suppression scale $M$. The lifetimes corresponding to suppression scales $M=10^{8},10^{12},10^{16}$, are, respectively,  $\log_{10}(\frac{\tau}{T_U})=-45,-90,-110$ while for the Standard Model $\log_{10}(\frac{\tau}{T_U})=540$.
\label{scaleplots2}
}
\end{figure}

In this section we will discuss how lowering the suppression scale $M$ in (2.1) changes our results.  To analyse this problem qualitatively it is enough to use the analytical approximation we presented in section \ref{anatyticalresults}~. When nonrenormalisable operators are positive, lowering the suppression scale $M$ corresponds simply to making the potential positive not far above $M$. The action (the exponent in \eqref{simpletime}) increases because the position of the minimum of $\lambda_{eff}$ shifts towards smaller energy scales and the value of $|\lambda_{eff}|$ decreases, which is shown in Figure~\ref{scaleplots1}.   
In the case with positive $\lambda_8$ and negative $\lambda_6$ this dependence is smaller as shown in Figure~\ref{scaleplots2} . The new minimum is deeper and changing the scale changes $\lambda_{eff}$ by a small fraction of its value  so the resulting lifetimes are much less scale dependent. In fact, in this case scale dependence of lifetime comes mostly from the prefactor in \eqref{simpletime},  because the size of the bounce is $\phi_0\approx \mu_{min}\propto M$. 
 
The last possibility is a potential unbounded from below which again corresponds to quickly decaying solutions, that depend on $M$ very much like in the previous case. Because their action is very close to zero, the actual dependence of the corresponding lifetime comes from the size of the bounce in the prefactor of \eqref{lifetime}. 

\section{Summary}
In this paper we have made a map of the vacua in the SM extended by nonrenormalisable scalar couplings, taking into account the running of the new couplings and going beyond the standard assumptions taken when calculating the lifetime of the metastable vacuum. Usually, one uses certain quasi-analytic approximations of the effective potential, \cite{WeinbergLee}, for the purpose of calculating the tunneling rate. In particular, one uses as the departure point the calculation for the quartic-like form of the effective potential at large field strenghts, while in the modified scalar potential it is the order 6 or order 8 coupling which naively dominates the potential in the large field domain. It is important to check the validity of such approximation and to search through a relatively wide scope of new couplings to find the actual behaviour of the scalar potential. The ultimate tool in this case is the direct numerical analysis, which however is not so straightforward because of the flatness of the effective potential. In this paper we present currently available results of such extended analysis of the modified SM scalar potential. For the purpose of the present study we have suppressed nonrenormalisable operators with derivatives. It should be noted that such operators can further destabilize the already unstable vacua, however complete study of this issue lies beyond the scope of this paper. 

It turns out that the simplified analytical approach represents reasonably well the actual numerical results. In general, we confirm that it is relatively easy to destabilise the  SM with the help of the Planck scale suppressed scalar operators. While it is possible to stabilise the SM  by adding such higher dimensional interactions and taking into account running of the new couplings, the generic effect is shortening the lifetime and hence further destabilisation of the SM electroweak vaccuum. This conclusion has been illustrated with the phase diagrams of modified SM-like models. It has been demonstrated that effective stabilisation can be achieved by lowering the suppression scale of higher order operators while picking up such combinations of new couplings, which do not deepen the new minima of the potential. Our results show the dependence of the lifetime of the electroweak minimum on the magnitude of the new couplings, including cases with very small couplings (which means very large effective suppression scale) and couplings vastly different in magnitude (which corresponds to two different suppression scales). 

\begin{center}
{\bf Acknowledgements}
\end{center}
 This work was supported by the Foundation for
Polish Science International PhD Projects Programme co-financed by the EU
European Regional Development Fund. This work has been supported by National Science Centre 
under  research grant DEC-2012/04/A/ST2/00099 and research grants DEC-2011/01/M/ST2/02466 and DEC- 2012/05/B/ST2/02597.

\end{document}